# Caldeira-Leggett oscillator bath simulation of the Prokof'ev-Stamp spin bath in strong coupling limits

Seyyed M.H. Halataei[1, 2, 3]

[1]*School of physics, Institute for Research in Fundamental Sciences (IPM)*
*P.O. Box 19395-5531, Tehran, Iran*
[2]*Pasargad Institute for Advanced Innovative Solutions (PIAIS), Tehran, Iran*
[3]*Department of Physics, University of Illinois at Urbana-Champaign,*
*1110 West Green Street, Urbana, Illinois 61801, USA*

Open quantum systems are subject to interaction with their surrounding environments. The environments are in general complex and intractable. At low temperatures, quantum environments are mapped onto two simpler universality classes of models, namely oscillator bath and spin bath models. The two models are commonly recognized as completely distinct at strong coupling limits. In particular, it is believed, that they cause two different dissipative relaxation rates when they act on qubits. It is also believed that at such limit the relaxation rate caused by the spin bath model cannot be simulated by the oscillator bath model. In this paper, I show, in contrast, that the oscillator bath model can simulate the effect of the spin bath model in strong coupling limit of the spin bath. I demonstrate that, by choosing right parameters for the oscillator bath model such that it induces the same rate and amount of bias energy fluctuations as those of the spin bath, the oscillator bath model can produce a relaxation rate equivalent to that of the spin bath model. This result implies that, as far as the relaxation rate is concerned, the distinction between the oscillator bath model and the spin bath model is far less than has been previously recognized.

# INTRODUCTION

Quantum theory was originally developed in the context of *isolated* microscopic systems whose interactions with their environments were negligible (e.g. the atoms in a beam). The theory was tested successfully in this domain in the early twentieth century and some of its founding fathers, such as Niels Bohr, believed that it would not be applicable in a larger domain where systems are strongly coupled to complex environments [1, 2].

Advancement of experimental techniques and equipments in recent decades, however, showed that quantum theory does apply in a broader range. It can well describe behaviors of *open* quantum systems, as large as a few microns, that are strongly coupled to their complex *environments* (e.g. the phase of the Cooper pairs, in SQUIDs, that is coupled to phonons, radiation field, normal electrons, nuclear spins, etc.) [3–10].

In handling complex environments coupled to open quantum systems, the *effects* that the environments exert on the principal systems are of the main interest, not the behaviors of the environments in their own right. As a result, theorists attempt to model complex environments by mapping them onto simpler ones that are better tractable and have the same *effects* on the principal systems. Two of these simple models that are well established in the literature are the Caldeira-Leggett oscillator bath and Prokof'ev-Stamp spin bath models [1, 3, 4, 11–19].

Leggett and his collaborators in a series of influential papers in 1980's, following Feynman and Vernon 1960's work [20], studied the oscillator bath model and showed that many quantum environments map onto this model [1, 3, 4, 21]. Almost a decade later, Prokofev and Stamp claimed that there is another type of quantum environment at low temperatures that cannot be mapped onto the oscillator bath model and has strikingly different effects on systems. This model was called the spin bath model. Prokofev and Stamp and their collaborators in a series of impactful papers in 1990's studied the spin bath model and their effect on quantum systems [11–17]. Both models have been being extensively used in the literature, however, they are commonly recognized as two distinct models[19, 22]. In this paper we show that there is a way to reconcile the two models and find an oscillator bath which simulates effects of the spin bath.

The *oscillator bath model* consists of a set of non-interacting simple harmonic oscillators that are individually coupled to the principal system (Fig. 1). Caldeira and Leggett [3, p. 439] showed that at absolute zero temperature, any arbitrary environment whose each degree of freedom is only weakly perturbed, by the principal system, can be mapped onto an oscillator bath [23].

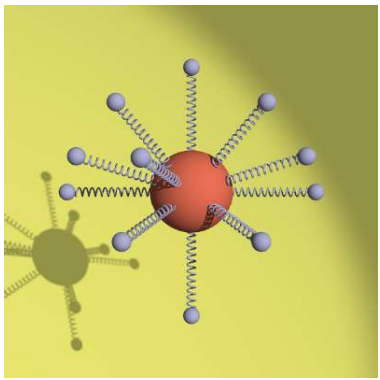

FIG. 1: Caldeira-Leggett oscillator bath model. The principal system is illustrated by the red ball. The environment are simulated by a set of simple harmonic oscillators, the gray balls, that are attached to the principal system and independently interact with it. Many environments at low temperatures can be mapped onto this model. Graphic is adapted from [24]
.

The oscillator bath model has been extensively used in the literature to model phonons, electrons, magnons, spinons, holons, quasiparticles, etc. at low energies and temperatures [4, 18, 19, 25–29].

The *spin bath model*, on the other hand, consists of microscopic spins that are independently coupled to the principal system (Fig. 2) . In real scenarios, these spins usually interact with one another weakly [15, 30].

The spin bath model usually represents the effects of nuclear spins, paramagnetic spins, and defects and can be studied at both weak and strong coupling limits [11, 13–17, 30–36].

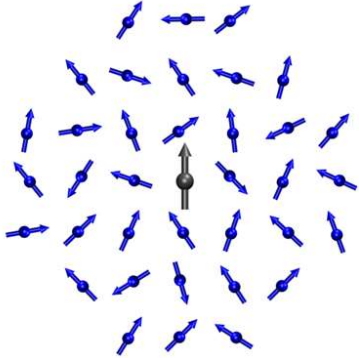

FIG. 2: Prokof'ev-Stamp spin bath model. The environment is simulated by a set of spins (the blue spins). The principal system is a two-level system which is illustrated by the gray spin. Many environments that are not trivially mapped onto the oscillator bath model, can be mapped onto the spin bath model at low temperatures. Graphic is adapted from [37]
.

It is intriguing to explore the possibility of simulating one of these models, i.e. the spin bath or oscillator bath, by the other one. If such simulation exists, then all ingredients of an environment at low temperatures, such as electrons, phonons, magnons, or nuclear spins, paramagnetic spins, defects, etc., can be described by one single model, say the oscillator bath model. This will simplify the study of evolution of an open quantum system in presence of an environment significantly.

Caldeira, Neto and de Carvalho [38] demonstrated that the effect of a non-interacting spin bath, *in the weak coupling limit*, on a principal system can be simulated by an oscillator bath whose spectral density function is suitably chosen [38].

Weiss [19] reached the same result by use of the fluctuation-dissipation theorem for each degree of freedom of the environment, which is permissible in the weak coupling limit (Sec. 3.5 and 6.1 of [19]).

Despite this success in simulating non-interacting spin bath in the weak coupling limit by an oscillator bath, the scheme was not extended to the case of self interacting spin bath, mainly due to the difficulty of calculating spin correlation function for each spin in a self interacting environment.

Prokofev and Stamp, who solved the spin bath problem directly and introduced the theory of spin bath for the first time [11–17], claimed that in the strong coupling limit the spin bath is unmappable onto the oscillator bath model and it exerts totally different effects on systems. They found that in this limit the effect of an interacting spin bath on an effectively two-state system is to relax the system incoherently. More precisely, Prokof'ev and Stamp obtained that under most conditions the principal system undergoes an incoherent relaxation with relaxation rate

$$\Gamma(\xi) \approx \frac{\Delta^2}{\xi_0} e^{-\left|\frac{\xi}{\xi_0}\right|}, \qquad (1)$$

where $\xi$ is the bias energy difference between the two minima of the double well energy landscape $E(\phi)$ of the principal system (Fig. 3), $\xi_0$ is the amount of fluctuation of the bias energy $\xi$ due to the presence of the bath, and $\Delta$ is the tunneling matrix element between the two localized states in either wells (Fig. 3).

Prokof'ev and Stamp, after arguing that under most conditions the relaxation rate of Eq. (1) (or in general Eq. (10)) would be obtained, claimed that this relaxation rate cannot be obtained from an oscillator bath [15]:

*"All of this is in complete contrast to how inelastic tunneling works in the presence of an oscillator bath; there the relaxation rate typically increases as one moves away from resonance"* (i.e. as the bias energy increases).

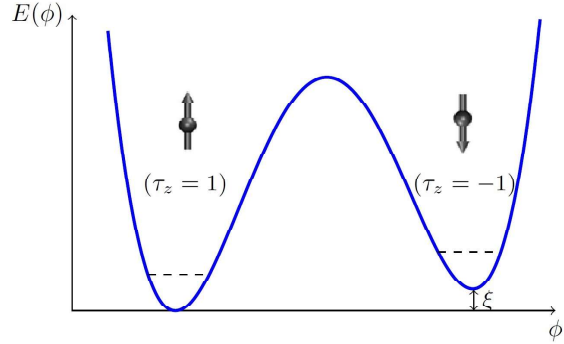

FIG. 3: Double well potential or energy landscape $E(\phi)$ [40] of the high energy Hamiltonian of the principal system. $\xi$ is the bias energy difference between the two minima [41] and $\phi$ can be multidimensional. The dashed lines represents the energy of the localized states in *absence* of tunneling.

They, hence, concluded that the spin bath model, in the strong coupling limit, has completely different effects on principal systems and cannot be simulated by the oscillator bath model [11, 13–17].

In this paper, we, however, show that the Caldeira-Leggett oscillator bath model *can* simulate the effect of Prokof'ev-Stamp spin bath model in the strong coupling limit of the spin bath. We demonstrate that an oscillator bath can cause an incoherent relaxation with a relaxation rate just like that of Eq. (1). We choose an oscillator bath that provides the same *rate* and *amount* of bias energy fluctuations as those of the spin bath and show that the oscillator bath causes the same relaxation rate as the spin bath does.

We emphasize that the effect we simulate in this paper is relaxation (not pure dephasing)! Relaxation is a dissipative process that usually transfers energy from the principal system to the environment, while pure dephasing is a non-dissipative process that conserves the energy of the system. In strong coupling limits, where most approximations fail[39], the problem of finding the relaxation rate for a system is non-trivial. A lot of efforts, hence, have been made in the literature to solve this problem even for the case of simple models such as the oscillator bath and spin bath (See e.g. Ref. [4] for the oscillator bath and Ref. [15] for the spin bath). On the other hand, the pure dephasing problem is quite easy and one can learn a complete solution of it in standard textbooks(See e.g. Ref. [18]). Here we tackle the relaxation problem. This can be easily verified from the spin bath Hamiltonian, Eq. (8), and its simulator oscillator bath Hamiltonian Eq. (12). None of those Hamiltonians preserve the populations of states, but they cause relaxation and dissipation. Putting it in the NMR language, what we do in this paper is mainly simulation of $T_1$ processes rather than only $\tau_\phi$ processes.

Finally, we note that the existence of the simulation we offer in this paper implies that for most practical purposes and as long as the effect of the environment in terms of its relaxation or decoherence rate is concerned, which is the case in practical applications of the spin bath [11], there is far less distinction between the spin bath and the oscillator bath models than has been previously recognized.

The organization of this paper is as follows: In the Results section we briefly review the spin bath and oscillator bath models, respectively, then we present the simulation of the spin bath model, in the strong coupling limit, by the oscillator bath model. We conclude and discuss the result in Discussion section.

## RESULTS

### Spin Bath Model

The spin bath model [11, 13–17] concerns the effect of the interaction of an effectively two-state system with an environment composed of microscopic spins, called spin bath (Fig. 2). The two-state system is usually a macroscopic system that at low temperatures can be treated as an effectively two-level system.

The spin bath model was mainly developed in studying the effect of environmental spins on molecule magnets at low temperatures. For the sake of illustration, we briefly introduce molecular magnets in this subsection and review the spin bath model in its context.

Molecule magnets compose a new class of magnetic materials that were discovered at the end of the twentieth century. Each molecule in these materials act like a *giant spin* when temperature is lowered sufficiently. The simplest model describing most of these single molecule magnets (SMMs) has a biaxial spin Hamiltonian as follows

$$H = -DS_z^2 + E(S_x^2 - S_y^2) \qquad (2)$$

where $0 < |E| < D$ and sign of $E$ can be positive or negative. In Hamiltonian (2), for $E > 0$, the easy axis is $z$, the medium axis is $y$, and the hard axis is $x$. Therefore, the Hamiltonian is called easy-plane-easy-axis Hamiltonian, referring to its easy plane, $YOZ$, and its easy axis, $X$ [30, 42].

Fe8 and Mn12 are the most investigated SMMs [30]. Fe8 molecules, for example, with formula $([(tacn)_6Fe_8O_2(OH)_{12}]^{8+})$, where *tacn* is the macrocyclic ligand $C_6H_{12}(NH)_3$, form a crystal (Fig. 4). In each molecule of Fe8 there are eight ions of $Fe^{3+}$ each with spin 5/2. These ions strongly interact with each other via intramolecule exchange interaction and below $400 mK$ they are locked into a fixed structure and act as a single giant spin with Hamiltonian (2) and total spin of $S = 10$. The values of $D$, $E$, for Fe8 along with three other SMMs are tabulated in Table I [42]. The chemical formulas of the other three SMMs can be found in Refs. [30, 35].

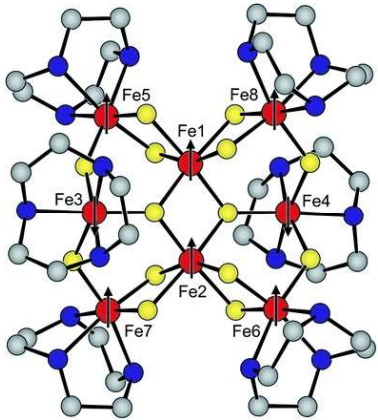

FIG. 4: Fe8 molecular structure. Graphic is adapted from [30].

TABLE I: Parameters of biaxial Hamiltonian (2) of two major single molecule magnets Fe8, Mn12 along with Mn4's [30, 35]

| SMM | $S$ | $D$ (K) | $|E|$ (K) |
|---|---|---|---|
| Fe8 | 10 | 0.295 | 0.056 |
| Mn12 | 10 | 0.65 | 0 |
| Mn4(S=9/2) | 9/2 | 0.68 | 0.064 |
| Mn4(S=8) | 8 | 0.43 | 0.029 |

At temperatures much smaller than $D$, such as $40\ mK$ as applied in many interesting experiments with SMMs, only the two lowest lying states are occupied. In such temperatures, the Hilbert space of Hamiltonian (2) can be truncated to the Hilbert space of an effective two-state system. The resultant Hamiltonian of this truncation procedure for Hamiltonian (2) is

$$H_S = -\frac{\Delta}{2}\hat{\tau}_x, \qquad (3)$$

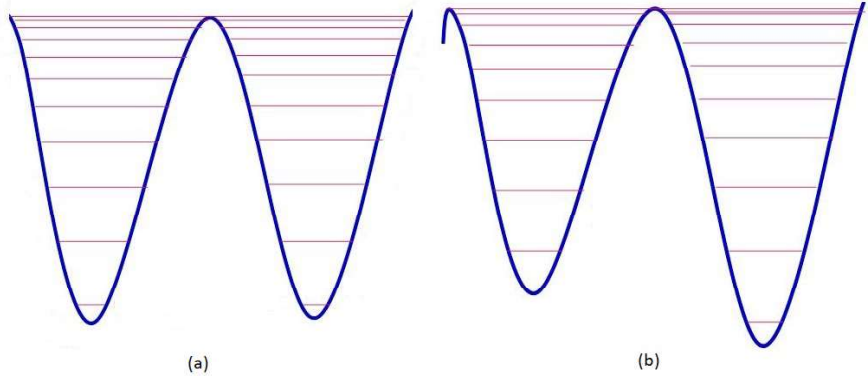

FIG. 5: Single Molecule Magnets form double well potentials with multiple energy levels in each well. Resonance occurs when one of the energy levels in one well lines up with another energy level of the other well. In (a) the wells are unbiased and the lowest level of the left well lines up with the lowest level of the right well, while in (b) the wells are biased and the bias energy $\xi$ is such that the lowest level of the left well coincides with the second lowest level of the right well. In both cases the system is at resonance. If quantum tunneling is also taken into consideration, the giant spin can tunnel from one well to the other well. This can only occur if the system is near resonance.

where we set $\hbar = 1$ as in the rest of this paper. Hamiltonian (2) has a double well energy landscape [40]. Thus, $\Delta$ is interpreted as the tunneling matrix element between the two localized states in either wells. One can bias the wells by applying a magnetic field along the $z$ axis in Hamiltonian (2). In such case, for sufficiently small biases, the effective two-state Hamiltonian becomes

$$H_S = -\frac{\Delta}{2}\hat{\tau}_x - \frac{\xi}{2}\hat{\tau}_z, \qquad (4)$$

where $\xi$ is the bias energy difference between the two minima[41] (Fig. 3). In above equations, $\hat{\tau}_x$ and $\hat{\tau}_z$ are Pauli matrices and we assumed that the localized states in either wells are eigenstates of $\hat{\tau}_z$. Thus, $\hat{\tau}_x$ takes the system from a localized state in one well to the other localized state in the other well (Fig. 3).

The energy levels of the lowest localized states in *absence* of tunneling are drawn by dashed lines in Fig. 3. The system is called to be at *resonance* when these levels coincide. The examples are $\xi = 0$ or when $\xi$ is such that the energy level of a higher state in one well lines up with that of the lowest state in the other well (Fig. 5).

In reality, however, the tunneling effect lifts the degeneracy such that near or at resonance the energy eigenstates become localized in both wells and the difference between energy eigenvalues becomes finite (Fig. 6)

To observe tunneling one usually applies magnetic field along the x or y axes [30]. The SMMs in their crystals interact with other SMMs and with microscopic spins such as nuclear spins, defects, phonons, etc.. Prokofev, Stamp and their collaborators (PS) developed the theory of spin bath, to solve the problem of the dynamics of giant spins in interaction with their surrounding *microscopic spins* [11–17].

The theory assumes a biaxial Hamiltonian for the giant spin (Eq. (1.3) of [14, p. 2903])

$$H_{PS} = \frac{1}{S}[-K_\parallel S_z^2 + K_\perp S_y^2] \qquad (5)$$

and adds to it the interaction Hamiltonian of the giant spin with microscopic spins (spin bath)

$$H_{int} = \frac{1}{2S}\sum_{k=1}^{N}\omega_k \ \vec{S}.\vec{\sigma}_k \qquad (6)$$

and also the self Hamiltonian of the spin bath

$$H_{B,\text{spin}} = \sum_{k,k'=1}^{N}\sum_{\alpha,\beta=1}^{3}V_{kk'}^{\alpha\beta}\hat{\sigma}_k^\alpha\hat{\sigma}_{k'}^\beta \qquad (7)$$

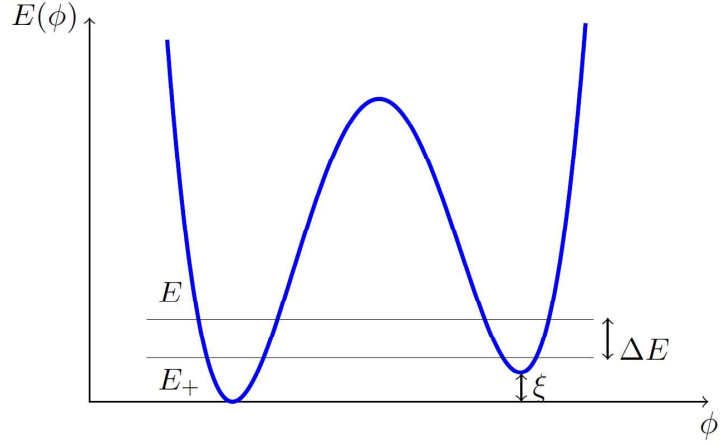

FIG. 6: Double well potential or energy landscape $E(\phi)$ of the principal system. The tunneling effect lifts the degeneracy in the energy levels. $E_\pm$ denote the ground state and first excited state energies of the system with splitting $\Delta E$. $\xi$ denotes the bias energy difference between the minima. For zero bias, $\xi = 0$, the energy splitting and the tunneling matrix element becomes equal, $\Delta E = \Delta \neq 0$. For $\xi \lesssim \Delta$ the eigenstates of the Hamiltonian are localized in both wells.

to form the total high energy Hamiltonian of the system plus environment (the spin bath). In Eqs. (6-7) $\hat{\vec{\sigma}}_k$ is the spin operator of the $k$th spin in the bath, $\omega_k$ is its coupling to the giant spin of the single molecule magnet, and $V_{kk'}^{\alpha\beta}$ are the interspin couplings of the spins of the spin bath.

Since in low temperatures only the ground state doublet of the giant spin is of the main importance, PS truncated the resultant Hamiltonian of the giant spin plus the spin bath to find the effective low energy Hamiltonian of the two-state system and the spin bath in the following form

$$H_{U,\text{spin}} = -\frac{\Delta}{2}\left\{\hat{\tau}_- \cos\left[\Phi - i\sum_{k=1}^{N}\vec{\alpha}_k \cdot \hat{\vec{\sigma}}_k\right] + \text{H.c.}\right\}$$
$$-\frac{\xi}{2}\hat{\tau}_z + \hat{\tau}_z \sum_{k=1}^{N}\vec{\omega}_k^{\parallel} \cdot \hat{\vec{\sigma}}_k + \sum_{k=1}^{N}\vec{\omega}_k^{\perp} \cdot \hat{\vec{\sigma}}_k$$
$$+\sum_{k,k'=1}^{N}\sum_{\alpha,\beta=1}^{3} V_{kk'}^{\alpha\beta}\hat{\sigma}_k^{\alpha}\hat{\sigma}_{k'}^{\beta}, \tag{8}$$

where $\hat{\vec{\tau}}$ are Pauli operators of the principal system, as before, and $\Phi$, $\vec{\alpha}_k$, $\vec{\omega}_k^{\parallel}$ and $\vec{\omega}_k^{\perp}$ are the parameters of the model that depend on the high energy Hamiltonian of the principal system (See Ref. [15]). The term in curly brackets in Eq. (8) is known as the topological term and in the absence of other terms it causes decoherence of the state of the principal system without dissipation of energy [43].

The tunneling matrix element $\Delta$ is typically the smallest energy scale in Hamiltonian (8). For Fe8, $\Delta \sim 0.1 \mu K$. The giant electronic spin, (S=10), of Fe8 molecule interacts with its surrounding nuclear spins and defects. These microscopic spins form the spin bath. The interspin couplings of spins in the spin bath are of order $V_{k,k'}^{\alpha\beta} \sim 1 \ \mu K$. Fe8 molecules form a crystal. In the crystal the giant spin of each Fe8 molecule interacts with that of the other Fe8 molecules in the crystal. The overall effect of other Fe8 molecules in the crystal is a quasi-static magnetic field which in terms of the bias energy is of order of $\xi \sim 0.1 \ mK$. The spin bath of each molecule, i.e. its nuclear spins and defects, causes the bias energy, however, to fluctuate by a root mean square value of $\xi_0 \sim 10 \ mK$. A similar hierarchy is valid for other typical SMMs. [11, 13–17, 30–33, 36].

In the strong coupling limit, when $V_{k,k'}^{\alpha\beta} < \vec{\omega}_k^{\parallel}, \vec{\omega}_k^{\perp} < \Delta$, the solution of Hamiltonian (8) under most conditions, and for practical cases (such as Fe8 and Mn12 [11]), is an incoherent relaxation of the principal system with relaxation rate, up to a factor of order unity [11, 15],

$$\Gamma(\xi) \approx \frac{\Delta^2}{\xi_0} e^{-\left|\frac{\xi}{\xi_0}\right|}. \tag{9}$$

The relaxation rate $\Gamma(\xi)$ is a decreasing function of $\xi$. As $\xi$ becomes significantly larger than $\xi_0$ the relaxation becomes significantly small and the system can hardly tunnel from one well to the other well. Thus, as long as the bias energy of the system $\xi$ is smaller than the width of the fluctuating field of the spin bath in energy unit $\xi_0$, the system will tunnel, otherwise it will remain in one well.

We note that in some cases of theoretical interest the relaxation rate is smaller than that of Eq. (9) and is given by

$$\Gamma(\xi) \approx \frac{\Delta^2}{\Gamma_2} e^{-\left|\frac{\xi}{\xi_0}\right|} \tag{10}$$

where $\Gamma_2 \geq \xi_0$ and depends on the parameters of the spin bath [11, 15]. This completes our review of the spin bath model. In the next subsection we turn to oscillator bath model and give a brief review of it. Then, in Sec. we replace the spin bath, for the problem of interaction of a qubit with an environment, with the oscillator bath and choose the parameters of the oscillator bath such that it simulates the effect of the spin bath and produces relaxation rates similar to that of Eqs. (9-10).

### Oscillator Bath Model

In this subsection we briefly review the Caldeira-Leggett oscillator bath model. The model studies the effect of an environment made of simple harmonic oscillators on a principal system (Fig. 1). The self Hamiltonian of the environment, which is called the oscillator bath, is

$$H_{B,osc} = \sum_i \frac{\hat{p}_i^2}{2m_i} + \frac{1}{2} m_i \omega_i^2 \hat{x}_i^2 \tag{11}$$

where $m_i$, $\omega_i$, $\hat{x}_i$, and $\hat{p}_i$, are mass, frequency, position and momentum operators of the $i$th oscillator in the bath, respectively. Many forms of interaction of the oscillators with a truncated two-state system can be cast into the following form [3, 4, 21], known as the *spin-boson* Hamiltonian,

$$H_{U,osc} = -\frac{\Delta}{2} \hat{\tau}_x - \frac{\xi}{2} \hat{\tau}_z + \frac{\hat{\tau}_z}{2} \sum_i c_i \hat{x}_i$$
$$+ \sum_i \frac{\hat{p}_i^2}{2m_i} + \frac{1}{2} m_i \omega_i^2 \hat{x}_i^2 \tag{12}$$

The dynamic of the principal system (the qubit) under influence of the above oscillator bath in various regimes of parameters has been investigated by Leggett, his collaborators, and many other authors in the literature [3, 4, 19, 21, 25–28]. In the regime that the bath has a continuum of low frequency modes, i.e. $\omega_i \lesssim \Delta$ for an infinite number of oscillators in the bath, the dynamic of the qubit is dissipative and irreversible. Dissipation causes transfer of energy from the qubit to the environment. That in turn gives rise to relaxation of the state of the qubit from excited states to the ground state ($T_1$ relaxation) as well as dephasing of the state of the qubit ($T_2$ relaxation).

We are interested in the solution of Eq. (12) in the small $\Delta$ limit when $\Delta \ll \xi$. This is the regime that we can relate to the spin bath model in the strong coupling limit since $\Delta$ is the smallest energy scale in the spin bath model in that limit, as discussed in the previous subsection. In the regime of $\Delta \ll \xi$, one can do perturbation theory in $\Delta$ to solve Eq. (12) for the dynamics of the qubit. This method is known as *golden* *rule* and is explained in details in Ref. [4]. Here we quote the result: The resultant dynamics of the qubit in this regime is an *incoherent relaxation* with relaxation rate

$$\Gamma(\xi) = \Delta^2 \int_0^\infty dt \; \cos(\xi\, t) \; \cos(\frac{Q_1(t)}{\pi}) \; e^{-Q_2(t)/\pi} \qquad (13)$$

where

$$Q_1(t) = \int_0^\infty \frac{J(\omega)}{\omega^2} \sin(\omega t) d\omega, \qquad (14)$$

$$Q_2(t) = \int_0^\infty \frac{J(\omega)}{\omega^2} (1 - \cos(\omega t)) \coth(\omega/2kT) \; d\omega, \qquad (15)$$

and

$$J(\omega) = \frac{\pi}{2} \sum_i \frac{c_i^2}{m_i \omega_i} \delta(\omega - \omega_i), \qquad (16)$$

is the spectral density function of the oscillator bath [3, 4].

In order to simulate the effect of the spin bath by the oscillator bath, we choose, in the next subsection, $J(\omega)$ and temperature of the oscillator bath such that it produces the same rate and amount of bias energy as the spin bath does. That, in turn, in the case we consider at least, induces the relaxation rates of the Prkof'ev-Stamp spin bath, Eq. (9-10).

## Simulation of Spin bath by oscillator bath in strong coupling limit

In this subsection we perform the simulation of the effect of the Prokof'ev-Stamp spin bath, in its strong coupling limit, by the Caldeira-Leggett oscillator bath. As we mentioned before in footnote [23], in the strong coupling limit some degrees of freedom of the environment are strongly perturbed, while in the weak coupling limit each degree of freedom of the environment is only weakly perturbed. This is with respect to the *environment*, not the *system* though. In both cases the principal system, here the effective two-state system, can be strongly perturbed. For example an oscillator bath with weak couplings $c_i$ in Eq. (12) can strongly perturb the qubit [4]. Therefore, with respect to the system there is no obvious difference between the weak coupling limit and the strong coupling limit. There may be a way to simulate the effect of an environment in the strong coupling limit through an environment in the weak coupling limit. Indeed, that is what we do in this subsection. We simulate the effect of the spin bath in the strong coupling limit, through an oscillator bath in the weak coupling limit.

In the weak coupling limit of the spin bath such a simulation has been already obtained as we mentioned in the introduction[19, 38]. What is remained is the demonstration of existence of such a simulation in the strong coupling limit of the spin bath. That is what we do here.

First we note, that an oscillator bath model can create a relaxation rate for a two-level system that decreases when the bias energy increases. This can be witnessed in Marcus's theory for charge transfer in solution, which can be cast in terms of a bath of oscillators coupled to a two-state Hamiltonian [44]. Marcus's theory predicts an *inverted regime* whose relaxation rate decreases with increasing bias. Therefore, the decrease of relaxation rate $\Gamma(\xi)$ with increase of the bias energy $\xi$ is not exclusive to the spin bath model. Nevertheless, it is interesting to see whether an oscillator bath can create a relaxation rate that is just like the one that a spin bath makes, i.e. Eqs. (9-10).

The effect of the Prokof'ev-Stamp spin bath on a two level system is a fluctuating bias energy (or equivalently a fluctuating magnetic field). Prokof'ev and Stamp argue that the *rate of fluctuation of the bias energy* and *the energy scale over which the bias energy fluctuates* are roughly equal (See p. 5795 of [11] and Sec. IV of [12]). Let us call this quantity $\xi_0$. On one hand, $\xi_0$ is the width of fluctuation of the bias energy and on the other hand is roughly equal to the rate of fluctuation of the bias energy in time.

An environment which has these two features in common with a spin bath may be able to simulate the effect of the spin bath. We see in this subsection that is indeed the case for an oscillator bath model we consider. We construct an oscillator bath model with the width of bias energy fluctuations roughly equal to $\xi_0$ and the rate of bias energy fluctuation also of order $\xi_0$. We show that such an oscillator bath mimics the effect of the spin bath and induces incoherent relaxation of the two-state system with relaxation rates according to Eqs. (9-10).

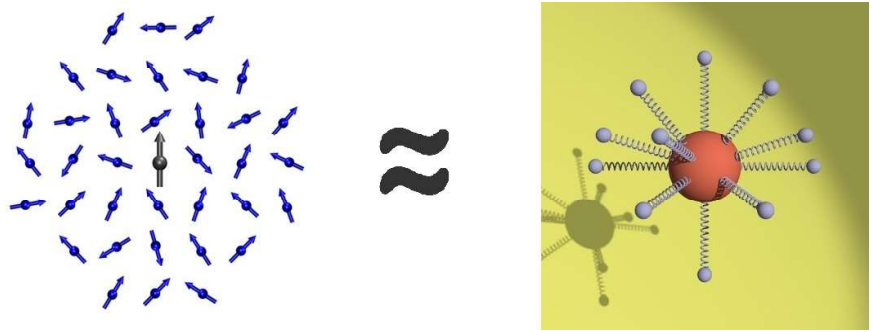

FIG. 7: As far as the relaxation rate is concerned, the Prokofev-Stamp spin bath model is equivalent to the Caldeira-Leggett oscillator bath model and can be simulated by that.

To this end, we choose the oscillator bath model of Eq. (12) with the same $\Delta$ and $\xi$ of the spin bath model of Eq. (8). Since $\Delta \ll \xi$ in the spin bath model, the oscillator bath model with such $\Delta$ and $\xi$ causes an incoherent relaxation, as discussed in the preceding subsection, with the relaxation rate of Eq. (13). We set the the temperature of the *simulator* oscillator bath to zero and choose the following spectral density function for it,

$$J(\omega) = 2\pi\alpha \; \omega \; e^{-\omega/\xi_0}, \tag{17}$$

where $\alpha = 1/2$.

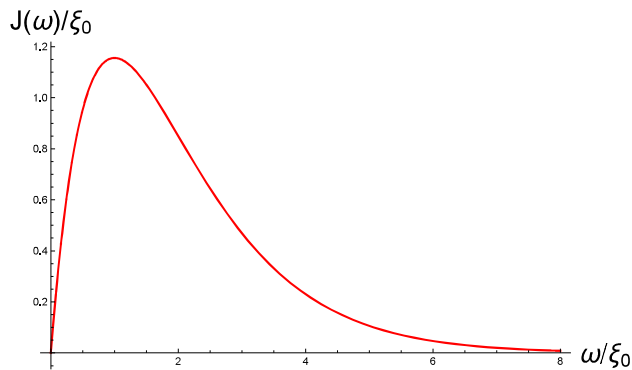

FIG. 8: Spectral density function $J(\omega)$ of the simulator oscillator bath of the spin bath, Eq. (17). $\xi_0$ is the *width* of fluctuation of the bias energy of the spin bath. The spread of $J(\omega)$ is also a few $\xi_0$. This, in turn, makes the *rate* of fluctuation of the bias energy to be of order of $\xi_0$.

The spectral density function of Eq. (17) is ohmic at small frequencies $\omega \ll \xi_0$, peaks at $\omega = \xi_0$ and then falls off and asymptotes to zero at large frequencies $\omega \gg \xi_0$ as shown in Fig. 8. The spread of the spectral density function of Eq. (17) is a few $\xi_0$. This is roughly equal to the inverse of the memory time of the noise that is produced by such an oscillator bath[45]. The inverse of memory time of the noise is, in turn, roughly equal to the *rate* of fluctuation of the bias energy caused by the noise. Therefore, the *rate* of bias energy fluctuations is of order of $\xi_0$. Thus, the spectral density function (17) gives the same *rate* of fluctuation of the bias energy as that of the spin bath.

The spectral density function (17) for an oscillator bath at zero temperature also produces the same *width* of the bias energy fluctuations that is produced by the spin bath. In order to show that, we first note that the spin-boson Hamiltonian (12) can be written as

$$H_{U,osc} = -\frac{\Delta}{2}\hat{\tau}_x + \frac{\hat{\xi}_B - \xi}{2}\hat{\tau}_z + H_{B,osc}$$

where

$$\hat{\xi}_B = \sum_i c_i \hat{x}_i \quad (18)$$

is the bias energy operator of the bath. The width of fluctuation of the bias energy is evidently $\sqrt{\langle \xi_B^2 \rangle}$. For an oscillator bath at zero temperature, one can estimate $\langle \xi_B^2 \rangle$ as follows

$$\langle \xi_B^2 \rangle = \langle \sum_{i,j} c_i c_j x_i x_j \rangle$$
$$\simeq \langle \sum_i c_i^2 x_i^2 \rangle \simeq \sum_i \frac{c_i^2}{m_i \omega_i} \quad (19)$$

where we neglected the cross terms $\langle x_i x_j \rangle$, used the ground state values for $\langle x_i^2 \rangle$, and set $\hbar = 1$ as before. The right hand side of (19) can be written in terms of the spectral density function (16),

$$\sum_i \frac{c_i^2}{m_i \omega_i} = \frac{2}{\pi} \int_0^\infty J(\omega) d\omega. \quad (20)$$

From (19) and (20) we deduce

$$\langle \xi_B^2 \rangle \simeq \frac{2}{\pi} \int_0^\infty J(\omega) d\omega. \quad (21)$$

We, hence, substitute our simulator bath spectral density function (17) into Eq. (21) to obtain

$$\langle \xi_B^2 \rangle = 2\xi_0^2. \quad (22)$$

Eq. (22) indicates that the *width* of the fluctuation of the bias energy in the spin bath model, $\xi_0$, is of the same order of magnitude as that of its simulator oscillator bath

$$\sqrt{\langle \xi_B^2 \rangle} \sim 1.4\ \xi_0. \quad (23)$$

Therefore, the two baths have this feature in common. They both have roughly the same *rate* and *amount* of bias energy fluctuations.

Given this fact, now let us show that the oscillator bath model of Eq. (12) with the spectral density function of Eq. (17) at zero temperature can simulate the effect of Prokofev-Stamp spin bath at strong coupling limit. That is it can produce the same relaxation rates of Eqs. (9-10) for the central two-state system.

To this end, we substitute the spectral density function (17) and zero temperature of the oscillator bath into Eqs. (14-15) to obtain

$$Q_1(t) = 2\pi\alpha\ \tan^{-1}\xi_0 t \quad (24)$$
$$Q_2(t) = \alpha\ \pi\ \ln(1 + \xi_0^2 t^2). \quad (25)$$

The above results for $Q_1(t)$ and $Q_2(t)$ are the same that are obtained in Ref. [4] for an ohmic oscillator bath with a cut off frequency when one sets the cutoff frequency to $\xi_0$ and temperature to zero (See Eq. (5.4) in Ref. [4]). We, next, substitute $Q_1(t)$ and $Q_2(t)$ from Eqs. (24-25), for the value of $\alpha = 1/2$, into Eq. (13) and take the integral. The integration is performed in Methods. The result is as follows

$$\Gamma(\xi) = \frac{\pi \Delta^2}{2\xi_0} e^{-\left|\frac{\xi}{\xi_0}\right|}. \tag{26}$$

This relaxation rate has the same form as that of the spin bath in Eq. (9), up to a factor of order unity.

To simulate the relaxation rate of Eq. (10) one needs to add a set of high frequency oscillators to the oscillator bath in Eq. (12). We add a set of oscillators with frequencies $\omega_i \gg \xi_0$ and spectral density function $J'(\omega)$ to the model of Eq. (12). The high frequency oscillators do not affect the dynamic of the qubit, however, they renormalize the tunneling matrix element to a smaller effective value $\Delta_{\text{eff}}$. This effect has been discussed in details in Sec. II of Ref. [4]. The qubit under the influence of the new bath, which is the previous bath plus a set of high frequency oscillators appended to it, with total spectral density function $J(\omega) + J'(\omega)$, relaxes incoherently, as before. However, the relaxation rate now becomes

$$\Gamma(\xi) = \frac{\pi \Delta_{\text{eff}}^2}{2\xi_0} e^{-\left|\frac{\xi}{\xi_0}\right|}. \tag{27}$$

One can choose the spectral density function of the high frequency oscillators $J'(\omega)$ such that

$$\frac{\Delta_{\text{eff}}^2}{\xi_0} = \frac{\Delta^2}{\Gamma_2}. \tag{28}$$

Thus, one obtains for the relaxation rate

$$\Gamma(\xi) = \frac{\pi \Delta^2}{2\Gamma_2} e^{-\left|\frac{\xi}{\xi_0}\right|} \tag{29}$$

that is the same as the relaxation rate of the spin bath in Eq. (10), up to a factor of order unity. One may fix the numerical factors by further tuning $J'(\omega)$ as well.

Relaxation rates of (26) and (29) are decreasing functions of the bias energy $\xi$ just like in the Prokof'ev-Stamp spin bath case. They also have the same form and match with those of the spin bath. Therefore, we deduce that the Caldeira-Leggett oscillator bath that we chose in this subsection can simulate the effect of the Prkof'ev-Stamp spin bath in the strong coupling limit of the spin bath. This is the result we wanted to obtain on this paper.

We note that our choice of the parameters of the simulator oscillator bath is not unique. One, for instance, may well find an oscillator bath at finite temperature that can give a similar effect. Our main demand from the simulator oscillator bath model was that it should provide the same rate and amount of fluctuation of the bias energy and we found that there is a model which fulfills these two demands and produces the same effect as that of the spin bath. It will be interesting to determine whether and why these two demands are sufficient to simulate the effect of the spin bath, which we leave for future work. It will also be intriguing to find how the result we obtained here fits in the broader picture of simulation of quantum noise by classical noise [46, 47], which we also leave for future work.

Moreover, we note that the two bath models are different in their own rights, but they exert the same effect on the system. In the spin bath model, the spins may interact with one another, while in the oscillator bath model, the oscillators are non-interacting. Each spin in the spin bath, individually, has a few number of excited states, while each oscillator in the oscillator bath has an infinite number of excited states available to it. However, given that studying the principal system, not the environment, is of main interest, as long as the effects of the two environments are similar, the existing differences between the two environments are not relevant. In other words, an environment is modeled only to study its effects on the system, not to understand the environment itself. We demonstrated that in spite of the differences between the spin bath and oscillator bath, the overall effect of the two baths on qubits can be similar in the strong coupling limit of the spin bath.

## DISCUSSION

We have shown in this paper that an oscillator bath can simulate the effect of the spin bath in the strong coupling limit of the spin bath. This is the limit that has been thought to have strikingly different effects on principal systems. We showed, however, that by choosing an oscillator bath that induces roughly the same rate and amount of the bias energy fluctuations as those of the spin bath, the oscillator bath can simulate the effect of the spin bath.

The result of this paper reconciles the two quantum environment models after three decades. It can affect the standard literature on quantum dissipative systems [19, 22] that, following Prokofev-Stamp work [11–17], recognized the view that the two models have strikingly different effects. Our result indicates that for most practical purposes the difference between the spin bath and the oscillator bath, as far as the relaxation rate is concerned, which is the case for most current practical applications, is far less than has been previously recognized.

In this paper, we have been more concerned with the results obtained by Prokof'ev-Stamp for effects of strongly coupled spin baths on qubits. Similar types of calculations have been done by Anupam Garg and collaborators in a series of papers (See for example [31, 32, 36] and references therein). They found qualitatively a similar effect. Quantitatively, however, they obtained that the relaxation rate decreases as $e^{-\xi^2/\xi_0^2}$ by the increase of the bias energy. It will be interesting to find the right parameters of an oscillator bath that can also quantitatively simulate the effect of the Garg spin bath. Our initial calculations show that an oscillator bath with a spectral density function that is highly peaked at low frequencies can simulate the Garg spin bath. We leave the details of calculations and simulation of the Garg spin bath for future works.

## METHODS

Here, we perform the integration in Eq. (13) for $Q_1(t)$ and $Q_2(t)$ of Eqs. (24-25) when $\alpha = 1/2$.

First, we substitute $Q_1(t)$ and $Q_2(t)$ from Eqs. (24-25) into cosine and exponential functions, respectively, to get,

$$\cos(Q_1(t)/\pi) = \frac{1}{\sqrt{1+\xi_0^2 t^2}}, \tag{30}$$

$$\exp(-Q_2(t)/\pi) = \frac{1}{\sqrt{1+\xi_0^2 t^2}}. \tag{31}$$

Next, we substitute the above results into Eq. (13)

$$\Gamma(\xi) = \Delta^2 \int_0^\infty \frac{\cos(\xi\, t)}{1+\xi_0^2 t^2}\, dt. \tag{32}$$

The integrand in Eq. (32) is an even function of $t$. Therefore, we can extend the integral to $-\infty$ for a prefactor of $1/2$,

$$\Gamma(\xi) = \frac{\Delta^2}{2} \int_{-\infty}^\infty \frac{\cos(\xi\, t)}{1+\xi_0^2 t^2}\, dt. \tag{33}$$

Now, we use the identity

$$\cos \xi t = \frac{e^{i|\xi|t} + e^{-i|\xi|t}}{2} \tag{34}$$

in Eq. (33) and split the integral into two parts to prepare them for contour integration

$$\Gamma(\xi) = \frac{\Delta^2}{4} \int_{-\infty}^\infty \frac{e^{i|\xi|t}}{1+\xi_0^2 t^2}\, dt + \frac{\Delta^2}{4} \int_{-\infty}^\infty \frac{e^{-i|\xi|t}}{1+\xi_0^2 t^2}\, dt \tag{35}$$

The above integrands have poles at $t = \pm i/\xi_0$. We close the contour of the first integral in the upper half plane and the second one in the lower half plane. Then we use the residue theorem to obtain

$$\begin{aligned}\Gamma(\xi) &= \frac{\Delta^2}{4}\left\{2\pi i \frac{e^{i|\xi|i/\xi_0}}{2i\xi_0} - 2\pi i \frac{e^{-i|\xi|(-i)/\xi_0}}{-2i\xi_0}\right\} \\ &= \frac{\pi\Delta^2}{2\xi_0}e^{-\left|\frac{\xi}{\xi_0}\right|}\end{aligned} \qquad (36)$$

Eq. (36) is the result we wanted to obtain and is the same as Eq. (26) in the text.

## ACKNOWLEDGEMENT

I am indebted to Anthony J. Leggett, my adviser during Ph.D., for stimulating discussions and fruitful comments. This work was motivated and inspired by an interesting discussion between him and P.C.E. Stamp at the end of the workshop on Large Scale Quantum Phenomena in Biological Systems at Galiano Island in June 2014 where different views were expressed about the relevance of the oscillator bath model in simulation of environments when interactions of systems and environments are strong. The video of the discussion is available in Ref. [48]. I am grateful to Mahdieh Piran for her support. I also acknowledge Pasargad Institute for Advanced Innovative Solutions (PIAIS) under supporting grant scheme (Project SG1RQT2103-01) for its support.

---